\documentclass[preprint,showpacs,preprintnumbers,amsmath,amssymb,superscriptaddress]{revtex4}
\usepackage{graphicx}% Include figure files
\usepackage{dcolumn}% Align table columns on decimal point
\usepackage{bm}% bold math
\usepackage{color}
\font\scripti=cmmi7
\font\scriptscripti=cmmi5
\def\sib#1{\setbox0 = \hbox{\scripti #1}
  \kern-.02em\copy0\kern-\wd0
  \kern.04em\box0} % script italic bold 
\def\ssib#1{\setbox0 = \hbox{\scriptscripti #1}
  \kern-.02em\copy0\kern-\wd0
  \kern.04em\box0} % scriptscript italic bold
\font\tenib=cmmib10 % italic bold for math
\skewchar\tenib='177 \skewchar\tenib='177 \skewchar\tenib='177
\def\pbold#1{\setbox0 = \hbox{$ #1 $}
  \kern-.022em\copy0\kern-\wd0
  \kern.011em\copy0\kern-\wd0
  \kern.011em\copy0\kern-\wd0
  \kern.011em\copy0\kern-\wd0
  \kern.011em\box0} % poorman's bold
\usepackage{graphicx}% Include figure files
\usepackage{dcolumn}% Align table columns on decimal point
\usepackage{bm}% bold math

\def\up{\uparrow}
\def\down{\downarrow}

\def\lesssim{\ \raise.3ex\hbox{$<$}\kern-0.8em\lower.7ex\hbox{$\sim$}\ }
\def\gesim{\ \raise.3ex\hbox{$>$}\kern-0.8em\lower.7ex\hbox{$\sim$}\ }

\begin{document}
\title{Spin-dipole mode in a trapped Fermi gas near unitarity}
\author{Hiroyuki Tajima}
\affiliation{Quantum Hadron Physics Laboratory, RIKEN Nishina Center (RNC), Wako, Saitama, 351-0198, Japan} 
\author{Alessio Recati}
\affiliation{INO-CNR BEC Center and Dipartimento di Fisica, Universit\`{a} di Trento, 38123 Povo, Italy}
\affiliation{Trento Institute for Fundamental Physics and Applications, INFN. 38123, Trento, Italy}
\author{Yoji Ohashi}
\affiliation{Department of Physics, Keio University, Hiyoshi, Kohoku-ku, Yokohama, 223-8522, Japan}
\date{\today}

\begin{abstract}
We theoretically investigate the spin-dipole oscillation of a strongly interacting Fermi gas in a harmonic trap. By using a combined diagrammatic strong-coupling theory with a local density approximation and a sum rule approach, we clarify the temperature dependence of the spin-dipole frequency near the unitarity, which is deeply related to the spin susceptibility, as well as pairing correlations.
While the spin-dipole frequency exactly coincides with the trap frequency in a non-interacting Fermi gas, it is shown to remarkably be enhanced in the superfluid state, because of the suppression of the spin degree of freedom due to the spin-singlet Cooper-pair formation.
In strongly interacting Fermi gases, this enhancement occurs even above the superfluid phase transition temperature, due to the strong pairing correlations. 
\end{abstract}
\maketitle
\section{Introduction}
Recently, ultracold atomic gases give us an ideal testing ground for the study of strongly correlated systems~ \cite{Bloch,Giorgini,Strinati}.
The most remarkable feature of this system is the controllability of the interatomic interaction associated with a Feshbach resonance~\cite{Chin}.
Indeed, by using this advantage various strong-coupling phenomena have been examined in strongly interacting Fermi gases.
\par
In particular, a unitary Fermi gas has attracted much attention because of its universal property, where the scattering length $a_{s}$ is adjusted to be divergent ($a_{s}=\pm \infty$)~\cite{Ho,Nascimbene,Horikoshi,Ku}.
Whereas the unitary Fermi gas does not depend on any scales associated with the interaction,
the presence of strong pairing fluctuations is anticipated.  
Indeed, the photoemission spectrum measurements near the superfluid critical temperature $T_{\rm c}$
indicate the existence of the pseudogap in single-particle excitations~\cite{Stewart,Gaebler,Perali,Sagi}, originating from strong pairing fluctuations (for a review, see Ref.~\cite{Mueller}).
On the other hand, it has also been shown that the observed equation of state in a unitary Fermi gas can be well reproduced by the Fermi liquid theory, without including strong pairing fluctuations \cite{Nascimbene2}. 
Thus, another quantity being sensitive to pairing fluctuations is needed, to understand such a strongly interacting system.
\par
Since the formation of singlet pairs suppresses the spin degrees of freedom, the spin susceptibility is a promising candidate for this purpose.
The so-called spin-gap phenomenon has been predicted~\cite{Enss,Palestini,Tajima1,Tajima2}, where the spin susceptibility is suppressed in the pseudogap regime.
While the spin susceptibility has recently been accessible experimentally in cold Fermi gas physics~\cite{Sanner,Sommer},
this many-body phenomenon has not been observed yet.
Furthermore, the spin-dipole frequency~\cite{Vichi,Recati}, which is also deeply related to the spin susceptibility, has experimentally been observed~\cite{Valtolina}.
Indeed, this quantity has successfully been used to observe the ferromagnetic behavior in the upper energy branch of a repulsive Fermi gas~\cite{Valtolina}.
The measurement of the spin-dipole frequency in the (attractive) lower energy branch is also reported in Ref.~\cite{Valtolina}.
While the spin-dipole frequency in a non-interacting Fermi gas is equal to the trap frequency due to Kohn's theorem with respect to the dipole mode even at finite temperatures~\cite{Kohn,OhashiK}, the large enhancement of this frequency has been observed in the unitary regime.
\par 
Another interesting aspect of the spin-dipole mode is an analogy with the giant dipole resonance (GDR) in nuclei~\cite{Lipparini}.
In such nuclear systems, the strong neutron-proton interaction plays an important role.
The excitation-energy dependence of GDR has been investigated to see effects of collective motions, as well as thermodynamic properties of excited nuclei~\cite{Santonocito}. 
\par
In this work, we discuss the spin-dipole frequency in an attractively interacting Fermi gas in a harmonic trap, by using a combined extended $T$-matrix approximation (ETMA)~\cite{Kashimura} with a local density approximation (LDA) and a sum rule approach.
Such a diagrammatic approach can, not only reproduce the observed spin susceptibility, but also connect the spin susceptibility with pairing-fluctuation corrections in a homogeneous two-component Fermi gas~\cite{Tajima1,Tajima2}.
Using this, we show effects of strong pairing interactions on the spin-dipole frequency in a strongly interacting trapped Fermi gas.
We also compare our numerical results with the recent experiment done near the unitarity limit.
\par
This paper is organized as follows.
In Sec.~\ref{sec2}, we explain our theoretical framework for the spin-dipole mode in a trapped Ferm gas.
In Sec.~\ref{sec3}, we discuss strong-coupling corrections on the spin-dipole frequency.
%Finally, we summarize our conclusions  in Sec.~\ref{sec4}.
In what follows, we take $\hbar=k_{\rm B}=1$.

\section{Formalism}
\label{sec2}
%In order for our paper to be self-contained, we begin by explaining the details of the combined ETMA and LDA formalism.
We start by considering a homogeneous three-dimensional two-component Fermi gas with a contact-type interaction. The Hamiltonian is given by
% \begin{eqnarray}
% \label{eq1}
% H&=&\sum_{\sigma=\up,\down}\sum_{\bm{p}}\xi_{\bm{p},\sigma}c_{\bm{p},\sigma}^{\dag}c_{\bm{p},\sigma}\cr
% &&-U\sum_{\bm{p},\bm{k},\bm{q}}c^{\dag}_{\bm{p}+\bm{q}/2,\up}
% c^{\dag}_{-\bm{p}+\bm{q}/2,\down}c_{-\bm{k}+\bm{q}/2,\down}
% c_{\bm{k}+\bm{q}/2,\up},
% \end{eqnarray}
\begin{eqnarray}
\label{eq1}
H&=&\sum_{\sigma=\up,\down}\sum_{\bm{p}}(\xi_{\bm{p}}-\sigma h)c_{\bm{p},\sigma}^{\dag}c_{\bm{p},\sigma}\cr
&&-U\sum_{\bm{p},\bm{k},\bm{q}}c^{\dag}_{\bm{p}+\bm{q}/2,\up}
c^{\dag}_{-\bm{p}+\bm{q}/2,\down}c_{-\bm{k}+\bm{q}/2,\down}
c_{\bm{k}+\bm{q}/2,\up},
\end{eqnarray}
where $c_{\bm{p},\sigma}$ ($c_{\bm{p},\sigma}^\dagger$) is the annihilation (creation) operator of a Fermi atom with momentum $\bm{p}$ and pseudospin $\sigma$, $\xi_{\bm{p}}=p^2/(2m)-\mu$ the kinetic energy ($m$ being the atomic mass) measured from the chemical potential $\mu$, and $h$ an effective magnetic field, i.e., (twice) the difference of the chemical potentials between the $\sigma=\uparrow$ and $\downarrow$ components. The interaction strength $-U$ is related to the $s$-wave scattering length $a_s$ as
\begin{eqnarray}
\label{eq2}
U=\left[\sum_{\bm{p}}^{|\bm{p}|\leq \rm p_{\rm c}}\frac{m}{p^2}-\frac{m}{4\pi a_s}\right]^{-1},
\end{eqnarray}
%At unitarity, i.e., $a_s\rightarrow\infty$, one can introduce a momentum cut-off
where $p_{\rm c}$ is the cut-off momentum.
In order to study the superfluid phase we introduce the superfluid order parameter $\Delta$, to rewrite the model Hamiltonian
in Eq. (\ref{eq1}) as~\cite{Takada1988},
\begin{eqnarray}
\label{eq3}
H=\sum_{\bm{p}}\psi^\dag_{\bm{p}}\left[\xi_{\bm{p}}\tau_3-h-\Delta\tau_1\right]\psi_{\bm{p}}
-U\sum_{\bm{q}}\rho_{\bm{q},+}\rho_{\bm{q},-},
\end{eqnarray}
where $\psi_{\bm{p}}=(c_{\bm{p},\up},c^{\dag}_{-\bm{p},\down})^{\rm t}$ is the two-component Nambu-field~\cite{Takada1988},
 $\tau_{i=0,1,2,3}$ the Pauli matrices and $\rho_{\bm{q},\pm}=\sum_{\bm{p}}\psi_{\bm{p}+\bm{q}/2}^{\dag}\tau_{\pm}\psi_{-\bm{p}+\bm{q}/2}$
are the generalized density operators with $\tau_{\pm}=(\tau_1\pm i\tau_{2})/2$.

As mentioned in the introduction, we include effects of a harmonic trap within LDA. 
In the present case of spin-independent trap potential $V(r)=\frac{1}{2}m\omega_{\rm tr}^2r^2$, in LDA the chemical potential is simply given by~\cite{Giorgini}
\begin{eqnarray}
\label{eq4}
\mu(r)=\mu-\frac{1}{2}m\omega_{\rm tr}^2r^2,
\end{eqnarray}
where $\omega_{\rm tr}$ is the trap frequency. Note that $h$ is position-independent in this case.
All the other quantities acquire spatial dependence via local solutions using  Eq.~(\ref{eq3}). 
Introducing $\xi_{\bm{p}}(r)=\xi_{\bm{p}}+m\omega_{\rm tr}^2r^2/2$, as well as the spatial dependent order parameter, $\Delta\rightarrow \Delta(r)$, we define the $2\times 2$ matrix LDA Green's function in the Nambu space as
\begin{eqnarray}
\label{eq5}
\hat{G}_{\bm{p}}(i\omega_n,r)^{-1}={i\omega_n-\xi_{\bm{p}}(r)\tau_3+h+\Delta(r)\tau_1-\hat{\Sigma}_{\bm{p}}(i\omega_n,r)},
\end{eqnarray}
where $\omega_n=(2n+1)\pi T$ ($n\in \mathbb{Z}$) is the fermion Matsubara frequency.
%%%%%%%%%%%%%%%%%%%%%%%%%%%%%%%%%%%%%%%%%%%%%%%%%%%%%%%%%%%%%%%%%%%%%%%%%%%%5
\begin{figure}[t]
\begin{center}
\includegraphics[width=7.5cm]{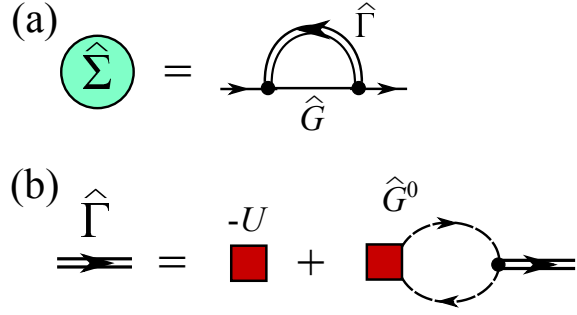}
\end{center}
\caption{Feynman diagram for the self-energy $\hat{\Sigma}$, as well as the many-body $T$-matrix $\hat{\Gamma}$ (double-solid lines).
The box represents a contact-type attractive interaction $-U$.
The solid and dashed lines show the dressed and bare Green's functions $\hat{G}$ and $\hat{G}^0$, respectively. 
}
\label{fig1}
\end{figure}
%%%%%%%%%%%%%%%%%%%%%%%%%%%%%%%%%%%%%%%%%%%%%%%%%%%%%%%%%%%%5
In ETMA, the LDA self-energy $\hat{\Sigma}_{\bm{p}}(i\omega_n,r)$ is diagrammatically described as Fig. \ref{fig1}(a), which gives
\begin{eqnarray}
\label{eq6}
\hat{\Sigma}_{\bm{p}}(i\omega_n,r)&=&-T\sum_{\bm{q},i\zeta_{n'}}\sum_{j,j'=\pm}\Gamma^{j,j'}_{\bm{q}}(i\zeta_{n'},r)\cr
&&\times\tau_j
\hat{G}_{\bm{p+\bm{q}}}(i\omega_n+i\zeta_{n'},r)\tau_{j'}
\end{eqnarray}
where $\zeta_{n'}=2n'\pi T$ ($n\in \mathbb{Z}$) is the boson Matsubara frequency.
The LDA many-body $T$-matrix $\hat{\Gamma}_{\bm{q}}(i\zeta_{n'},r)$, diagrammatically shown in Fig. \ref{fig1}(b), has the form,
\begin{eqnarray}
\label{eq7}
\hat{\Gamma}_{\bm{q}}(i\zeta_{n'},r)=-\left[1+U\hat{\Pi}_{\bm{q}}(i\zeta_{n'},r)\right]^{-1}U.
\end{eqnarray}
Here,  
\begin{eqnarray}
\label{eq8}
\left[\hat{\Pi}_{\bm{q}}(i\zeta_{n'},r)\right]_{j,j'}&=&T\sum_{\bm{p},i\omega_n}{\rm tr}\left[\hat{G}^0_{\bm{p}+\bm{q}}(i\omega_n+i\zeta_{n'},r)\right.\cr
&&\times\left.
\tau_{j'}\hat{G}^0_{\bm{p}}(i\omega_n,r)\tau_j\right],
\end{eqnarray}
is the LDA pair-correlation matrix,
where $\hat{G}^0_{\bm{p}}(i\omega_n,r)=\left[i\omega_n-\xi_{\bm{p}}(r)\tau_3+\Delta(r)\tau_1\right]^{-1}$ is the bare BCS Green's function~\cite{Strinati}.
\par
The local density $n_{\sigma}(r,T)$ is obtained from $\hat{G}_{\bm{p}}(i\omega_n)$ as
\begin{eqnarray}
\label{eq9}
n_{\up}(r,T)&=&T\sum_{\bm{p},i\omega_n}\left[\hat{G}_{\bm{p}}(i\omega_n,r)\right]_{11}e^{i\omega_n \delta}, \\ \nonumber
n_{\down}(r,T)&=&T\sum_{\bm{p},i\omega_n}\left[\hat{G}_{\bm{p}}(i\omega_n,r)\right]_{22}e^{-i\omega_n \delta},
\end{eqnarray}
where $\delta$ is an infinitesimally small positive number.
The Fermi chemical potential $\mu$ is determined by solving numerically the particle-number equation $N=N_{\uparrow}+N_{\downarrow}$
where 
\begin{eqnarray}
\label{eq10}
N_{\sigma}=\int d^3\bm{r}n_{\sigma}(r,T).
\end{eqnarray}
%In particular $N_{\uparrow}+N_{\downarrow}$ fixes the chemical potential $\mu$ (the magnetic field $h$).
\par
In the superfluid phase, we also determine the LDA superfluid order parameter $\Delta(r)$ from the gapless condition of the Nambu-Goldstone mode~\cite{OhashiGriffin},
\begin{eqnarray}
\label{eqSFO}
{\rm det}\left[1+U\hat{\Pi}_{\bm{q}=0}(i\nu_{n'}=0,r)\right]=0.
\end{eqnarray}
We define the LDA superfluid critical temperature $T_{\rm c}$ as the temperature below which $\Delta(r=0)$ becomes non-zero.

In the present work, we use the ETMA+LDA to estimate the spin-dipole mode frequency,
i.e., the frequency of the out-of-phase in-trap dipole motion of the two spin components. A rigorous upper bound~\cite{Recati} is given by the ratio between the energy weighted $m_1$ and the inverse energy weighted $m_{-1}$ sum rules for the spin-dipole operator $\sum_i z_{i,\uparrow}-\sum_i z_{i,\downarrow}$, where the sums run over all the $\uparrow$ and $\downarrow$ atoms, respectively.

While $m_1\propto N/m$, the sum rule $m_{-1}$ directly depends on the magnetic susceptibility of the gas $\chi(r,T)$, which can be calculated as, in LDA,
\begin{eqnarray}
\label{eq11}
\chi(r,T)=\lim_{h\rightarrow 0}\frac{n_{\up}(r,T)-n_{\down}(r,T)}{h}.
\end{eqnarray}
Eventually, the spin-dipole frequency $\omega_{\rm SD}$ is evaluated as \cite{Recati}
\begin{eqnarray}
\label{eq12}
\omega_{\rm SD}^2\le\frac{m_1}{m_{-1}}=\frac{N}{m\int d^3\bm{r}z^2\chi(r,T)}.
\end{eqnarray}
In this work, we numerically evaluate Eq.~(\ref{eq11}) with a small magnetic field $h=10^{-2}\varepsilon_{\rm F}$.
While Eq.~(\ref{eq12}) generally represents an upper bound, it is expected to give a very accurate estimation for the spin-dipole frequency, 
since the spin-dipole operator excites mainly a single mode. 
We briefly note that, at low frequency, a better estimation can 
be obtained by including the mass normalization~\cite{Recati} in the f-sum rule as $m_1\propto N/m^*$, which is, however, outside the scope of the present work,
and which is a higher-order effect on the spin susceptibility along the temperature evolution. 

\section{Results}
\label{sec3}

\begin{figure}[t]
\begin{center}
\includegraphics[width=8cm]{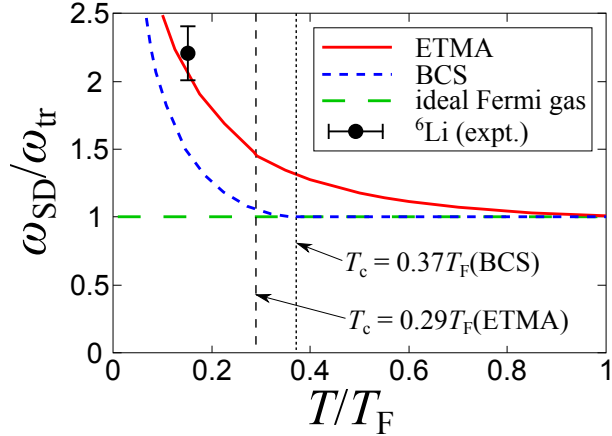}
\end{center}
\caption{Calculated spin-dipole frequency in a trapped unitary Fermi gas at finite temperature.
The solid, dotted, and dashed lines represent the results of ETMA, BCS, and an ideal Fermi gas.
The LDA critical temperatures of ETMA ($T_{\rm c}=0.29T_{\rm F}$) and BCS ($T_{\rm c}=0.37T_{\rm F}$) are shown, where $T_{\rm F}=(3N)^{1\over 3}\omega_{\rm tr}$ and $\omega_{\rm tr}$ are the Fermi temperature in a two-component ideal gas and the trap frequency, respectively.
The black circle shows the recent experimental result on a $^6$Li unitary Fermi gas~\cite{Valtolina}.
}
\label{fig2}
\end{figure}
Figure \ref{fig2} shows the temperature dependence of the spin-dipole frequency $\omega_{\rm SD}$ for a unitary Fermi gas in a harmonic trap.
The solid and short-dashed lines represent the results of ETMA and the BCS mean-field approximation (hereinafter, referred to as BCS), respectively. 
The BCS result is obtained by solving Eq.~(\ref{eq10}) without the self-energy correction, namely, $\hat{\Sigma}_{\bm{p}}(i\omega_n,r)=0$.
Our result shows an excellent agreement with the recent experiment at $T=0.151T_{\rm F}$~\cite{Valtolina}, with $T_{\rm F}=(3N)^{1\over 3}\omega_{\rm tr}$ being the Fermi temperature in an ideal two-component Fermi gas.
Although the experimental result was obtained in the condition that the two gas clouds of up and down spins are initially not fully overlapping,  ETMA  can quantitatively explain the magnetic properties of a unitary Fermi gas.
A similar agreement about the strong coupling corrections to the spin susceptibility have also been pointed out in Ref.~\cite{Tajima1,Tajima2}.
\par
As shown in Eq.~(\ref{eq12}),
$\omega_{\rm SD}^2$ is inversely proportional to the {\textsl{second moment}} of the local spin susceptibility $\chi(r,T)$.
Thus, $\omega_{\rm SD}$ becomes larger, for smaller $\chi(r,T)$.
\begin{figure}[t]
\begin{center}
\includegraphics[width=7cm]{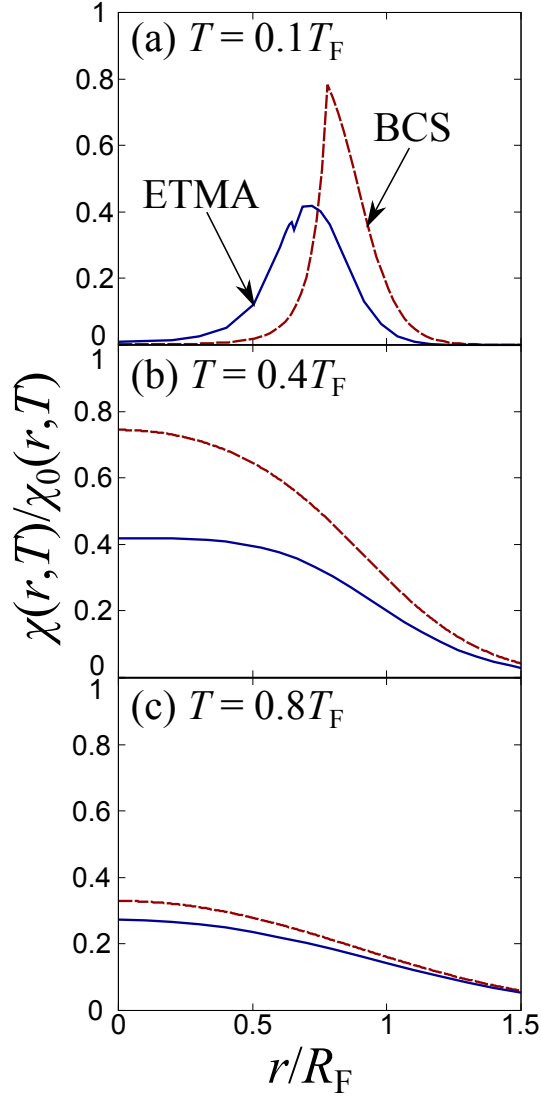}
\end{center}
\caption{Local spin susceptibility at (a) $T=0.1T_{\rm F}$ (superfluid phase), (b) $T=0.4T_{\rm F}$, and (c) $T=0.8T_{\rm F}$ in a trapped unitary Fermi gas. 
The solid and dashed lines represent the result of ETMA and BCS, respectively.
$\chi_0(r,T)$ is the Pauli susceptibility for homogeneous gases with number density $n_{\sigma}(r,T)$ (see text).
}
\label{fig3}
\end{figure}
We obtain a better insight into the large enhancement of $\omega_{\rm SD}$, from the spatial and temperature dependence of the local spin susceptibility $\chi(r,T)$. Figure~\ref{fig3} shows $\chi(r,T)$ as a function of $r$ for different temperatures. We introduce the ideal Thomas-Fermi radius $R_{\rm F}=\sqrt{2\varepsilon_{\rm F}/(m\omega_{\rm tr}^2)}$ with $\varepsilon_{\rm F}$ is the Fermi energy in an ideal two-component Fermi gas at $T=0$, as well as the Pauli susceptibility for a homogeneous gas,
$\chi_0(r,T)=(3m/2)[n_{\up}(r,T)+n_{\down}(r,T)]^{1\over 3}/(3\pi^2)^{2\over 3}$~\cite{Tajima3}. 
In the superfluid phase at $T=0.1T_{\rm F}$ shown in Fig.~\ref{fig3}(a), both the ETMA and BCS results exhibit peak structures.
In our LDA formalism, the system forms a shell structure with a superfluid core ($\Delta(r)\neq 0$), where the inside (outside) region is the superfluid (normal) phase.
In the superfluid region, the local spin susceptibility is largely suppressed due to the formation of singlet Cooper pairs.
While the ETMA result shows a larger $\chi(r,T)$ than that of BCS in the superfluid region,
the opposite occurs in the normal region.
This behavior reflects pairing fluctuations in each region.
In particular, in the normal phase at $T=0.4T_{\rm F}$ and $T=0.8T_{\rm F}$ shown in Fig.~\ref{fig3}(b,c), the ETMA result is always smaller than the BCS one. 
This difference originates from the formation of preformed Cooper pairs near $T_{\rm c}$ and the interaction effect becomes smaller at high temperature regime.
\par
We note that the calculation of BCS above $T_{\rm c}$ ($=0.37T_{\rm F}$) is equivalent to the non-interacting case.
Regarding this, $\omega_{\rm SD}$ is always equal to $\omega_{\rm tr}$ above $T_{\rm c}$ in the mean-field approximation.
However, as shown in Fig.~\ref{fig3} (b,c), $\chi(r,T)$ in the normal phase clearly has a temperature dependence even in the mean-field calculation.
Although effects of pairing fluctuations on the trap-averaged spin susceptibility is unclear due to the fact that it involves, not only pairing-correlations, but also temperature-dependent density profile~\cite{Tajima3},
the spin-dipole frequency is not affected by the latter effect.
\par
\begin{figure}[t]
\begin{center}
\includegraphics[width=7.5cm]{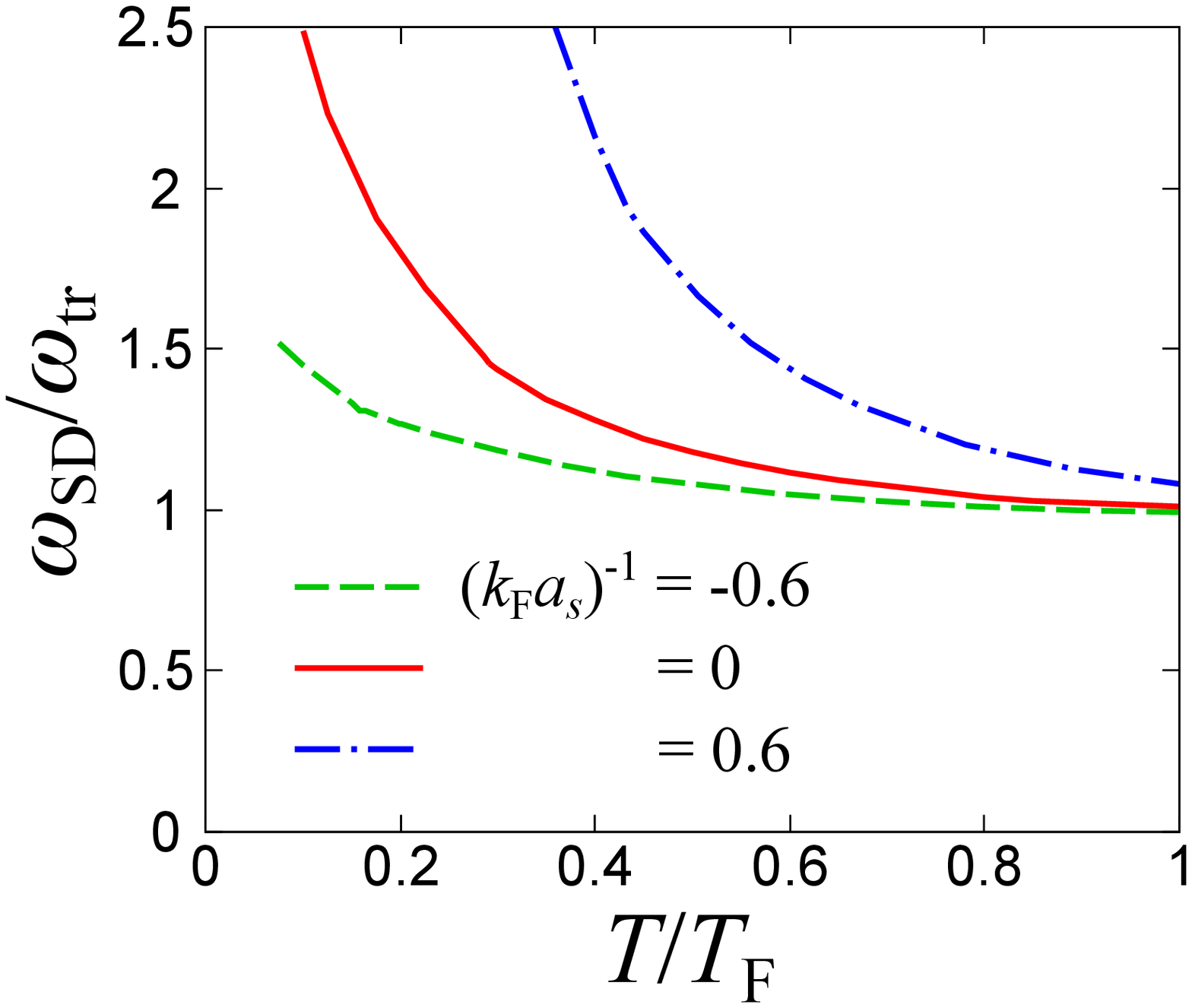}
\end{center}
\caption{Comparison of the spin-dipole frequencies at $(k_{\rm F}a_s)^{-1}=-0.6$ (dashed line),
$0$ (solid line) and $0.6$ (dot-dashed line).
$k_{\rm F}=\sqrt{2m\omega_{\rm tr}(3N)^{1\over 3}}$ is the LDA Fermi momentum.
}
\label{fig4}
\end{figure}
In Figure~\ref{fig4}, we report the temperature dependence of the spin-dipole frequency away from the unitarity.
In the high-temperature limit, the expected result $\omega_{\rm SD}=\omega_{\rm tr}$ is recovered, irrespective of the interaction strength.
On the other hand, $\omega_{\rm SD}$ diverges in the low-temperature limit for non-zero pairing interaction, where the spin susceptibility vanishes because of the singlet Cooper pairing. %\footnote{we show the eye-guide for the low-temperature result of $(k_{\rm F}a_s)^{-1}=-0.6$, because of the computational problem}.
Since the spin susceptibility becomes smaller for a weaker pairing interaction,
$\omega_{\rm SD}$ increases with increasing $(k_{\rm F}a_s)^{-1}$.
In particular, $\omega_{\rm SD}$ becomes substantially large in the strong-coupling regime ($(k_{\rm F}a_s)^{-1}\gesim 0.5$) where atoms form tightly bound molecules.
However, even in such a regime, we obtain $\omega_{\rm SD}=\omega_{\rm tr}$ 
for $T\gg E_{\rm b}$, with $E_{\rm b}=1/(ma_s^2)$ the two-body binding energy at $r=0$.
This indicates that the spin-dipole frequency is equal to the trap frequency where pairing correlations are negligible.
In the absence of the interaction effect, since each spin component exhibits a dipole oscillation independently, the spin-dipole frequency coincides with the dipole frequency, as well as therefore the trap frequency, due to Kohn's theorem~\cite{Kohn,OhashiK}, which exactly proves that the dipole frequency in trapped gases is always equal to the trap frequency.

In the strong-coupling high-temperature regime, the system can simply be described by a classical atom-molecule mixture~\cite{Tajima3}, so that the local spin susceptibility $\chi_{\rm cl}(r,T)$ is analytically given by
\begin{eqnarray}
\label{eqxcl}
\chi_{\rm cl}(r,T)=\frac{2\lambda}{T}\left(\frac{mT}{2\pi}\right)^{\frac{3}{2}}
\exp\left(-\frac{m\omega_{\rm tr}^2r^2}{2T}\right),
\end{eqnarray}
where $\lambda=e^{\mu/T}$ is the fugacity.
Such a classical mixture model is equivalent to the so-called Saha-Langmuir equation~\cite{Saha,Langmuir}.
Recently, the pair fraction predicted by the Saha-Langmuir equation shows good agreement with the cold atom experiment in the BEC side~\cite{Paintner}.
Substituting Eq. (\ref{eqxcl}) into Eq. (\ref{eq12}), one obtains
\begin{eqnarray}
\label{eqvsdcl}
\omega_{\rm SD}^{\rm cl}=\omega_{\rm tr}\sqrt{\frac{1}{6\lambda}\left(\frac{T_{\rm F}}{T}\right)^3}.
\end{eqnarray} 
Since $\lambda=(T_{\rm F}/T)^3/6$ in trapped ideal two-component gases, 
Eq. (\ref{eqvsdcl}) is consistent with Kohn's theorem ($\omega_{\rm SD}^{\rm cl}=\omega_{\rm tr}$).
In the presence of the molecular bound state, $\lambda$ is given by
\begin{eqnarray}
\label{eqlam}
\lambda=\frac{\sqrt{1+\frac{2}{3}\left(\frac{T_{\rm F}}{T}\right)^3\exp(E_{\rm b}/T)}-1}{2\exp(
E_{\rm b}/T)}.
\end{eqnarray}
Using this, one can analytically obtain the spin-dipole frequency in the strong-coupling high-temperature limit,
\begin{eqnarray}
\label{eqvsdcl2}
\omega_{\rm SD}^{\rm cl}=\omega_{\rm tr}\sqrt{\frac{\left(\frac{T_{\rm F}}{T}\right)^3\exp(E_{\rm b}/T)}{3\sqrt{1+\frac{2}{3}\left(\frac{T_{\rm F}}{T}\right)^3\exp(E_{\rm b}/T)}-3}}.
\end{eqnarray}
In this way, one can understand that the enhancement of $\omega_{\rm SD}$ in the strong-coupling regime due to the appearance of tightly bound molecules.
Indeed, Eq. (\ref{eqvsdcl2}) coincides with Eq. (\ref{eqvsdcl}) when $T\gg E_{\rm b}$ because molecules are thermally dissociated.
This result is in sharp contrast to the dipole mode which does not depend on the temperature and the interaction strength due to Kohn's theorem.

\begin{figure}[t]
\begin{center}
\includegraphics[width=7.5cm]{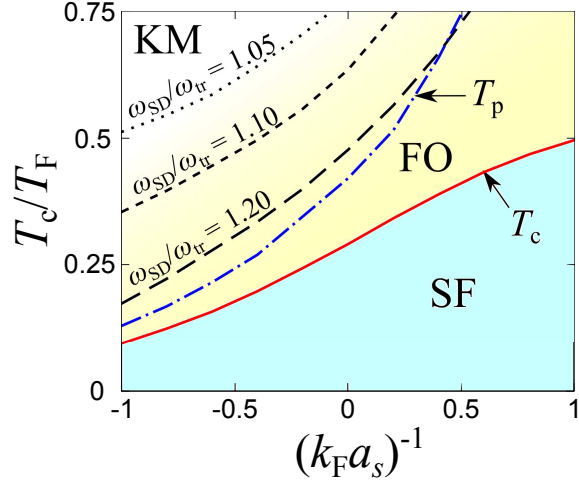}
\end{center}
\caption{Phase diagram of an attractively interacting trapped Fermi gas.
The solid line shows the LDA superfluid critical temperature $T_{\rm c}$, below which the system undergoes the superfluid phase (``SF").
While the spin-dipole frequency can be explained by the Kohn (dipole) mode in the high-temperature region (``KM"), it is strongly enhanced at low temperatures (fast oscillation region, ``FO").
Although there are no clear boundary between ``KM" and ``FO",
the crossover between these two regimes can be
characterized by the temperatures where $\omega_{\rm SD}/\omega_{\rm tr.}=1.05$ (dotted line), $1.10$ (dashed line), and $1.20$ (long-dashed line).
For comparison, we also plot the peak temperature $T_{\rm p}$ of the trap-averaged spin susceptibility shown in~\cite{Tajima3}.
}
\label{fig5}
\end{figure}

We summarize the phase diagram of an attractively interacting trapped Fermi gas from the viewpoint of the spin-dipole mode in Fig.~\ref{fig5}.
In this figure, we plot the temperatures where $\omega_{\rm SD}/\omega_{\rm tr}=1.05$, $1.10$, and $1.20$.
These three characteristic temperatures monotonically increase with increasing the pairing interaction strength.
One can find the smooth crossover from the high-temperature region (``KM"), where the spin-dipole oscillation can be explained by the two independent Kohn (dipole) modes of the spin $\sigma=\uparrow$ and $\downarrow $ gas clouds, to the fast-oscillation region (``FO") where $\omega_{\rm SD}$ largely deviates from $\omega_{\rm tr}$ due to the strong attractive interaction.
Although there is no clear phase boundary between KM and FO, 
this result indicates that $\omega_{\rm SD}$ is sensitive to the pairing interaction, as well as resulting singlet-pair formations. 
%In connection with nuclear physics,
%the $^3S_1$ neutron-proton interaction with the positive scattering length $a_{\rm np}$ plays a crucial role in the giant dipole resonance, where neutrons and protons oscillate in the out-of-phase due to an external field.
%In this case, the ``isospin" susceptibility, corresponding to the linear coefficient of the symmetry energy, is suppressed because of the isospin-singlet deuteron formation.
%Such a case correpsonds to the positive-scattering region in Fig.~\ref{fig5}. 
For reference, we also show in Fig.~\ref{fig5} the peak temperature $T_{\rm p}$ of the trap-averaged spin susceptibility in an attractive Fermi gas in a harmonic trap~\cite{Tajima3}.
$T_{\rm p}$ is close to the temperature where $\omega_{\rm SD}/\omega_{\rm tr}=1.20$,
in the entire crossover region.
\begin{figure}[t]
\begin{center}
\includegraphics[width=7.5cm]{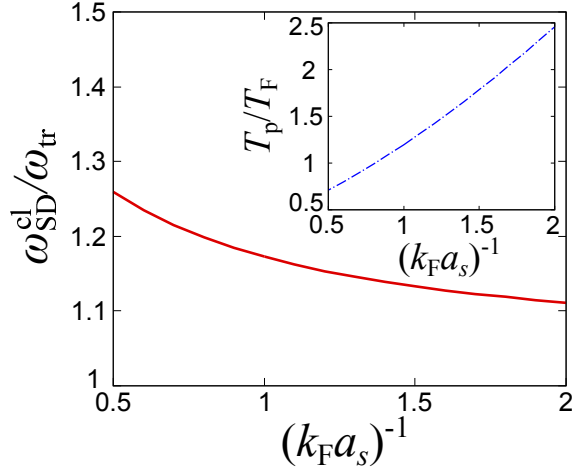}
\end{center}
\caption{Calculated spin-dipole frequency $\omega_{\rm SD}^{\rm cl}$ in the classical atom-molecule mixture at $T=T_{\rm p}$.
The inset shows the peak temperature $T_{\rm p}$ of the trap-averaged spin susceptibility~\cite{Tajima3}, as a function of $(k_{\rm F}a_s)^{-1}$, in the strong-coupling limit.
}
\label{fig6}
\end{figure}
\par
Figure~\ref{fig6} shows $\omega_{\rm SD}^{\rm cl}$ at $T=T_{\rm p}$ in the strong-coupling regime.
$T_{\rm p}$ as a function of $(k_{\rm F}a_s)^{-1}$ is also shown in the inset of Fig.~\ref{fig6}.
While $T_{\rm p}$ increases and $\omega_{\rm SD}^{\rm cl}$ at $T=T_{\rm p}$ slightly decreases with increasing the binding energy,
this indicates that the spin-dipole frequency starts to be enhanced when the trap-averaged spin susceptibility is suppressed by the singlet-pair formation.
\par
\begin{figure}[t]
\begin{center}
\includegraphics[width=7.5cm]{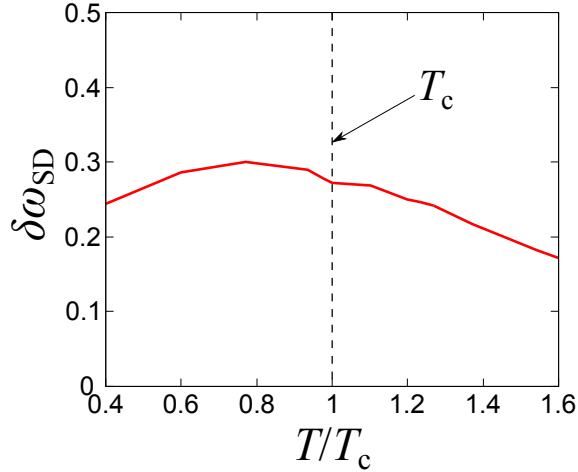}
\end{center}
\caption{Frequency shift $\delta\omega_{\rm SD}=\left(\omega_{\rm SD}^{\rm ETMA}-\omega_{\rm SD}^{\rm BCS}\right)/\omega_{\rm SD}^{\rm ETMA}$ due to the pairing correlations beyond the mean-field approximation near $T=T_{\rm c}$.
}
\label{fig7}
\end{figure}

To see how the emergent superfluid order and pairing fluctuations affect the spin-dipole mode below $T_{\rm c}$,
we plot in Fig.~\ref{fig7} the frequency shift $\delta \omega_{\rm SD}=\left(\omega_{\rm SD}^{\rm ETMA}-\omega_{\rm SD}^{\rm BCS}\right)/\omega_{\rm SD}^{\rm ETMA}$ near $T_{\rm c}$, where $\omega_{\rm SD}^{\rm ETMA(BCS)}$ is the spin-dipole frequency obtained by ETMA (BCS).
By definition, this quantity purely originates from the pairing correlations beyond the mean-field level.
$\delta \omega_{\rm SD}$ gradually increases with decreasing temperature in the high-temperature region,
and starts to decrease around $T=0.77T_{\rm c}= 0.23T_{\rm F}$.
The pairing effect on $\omega_{\rm SD}$ is most visible near this temperature.
Indeed, pairing fluctuations become strong near $T_{\rm c}$ in single-particle excitations~\cite{Perali2,Tsuchiya,Tsuchiya2,Ota}.
%It is in contrast to other characteristic temperatures for superfluid precursors such as pseudogap temperature~\cite{Tsuchiya,Tsuchiya2}, which can be seen above $T_{\rm c}$ as usual. 
Finally, in the low-temperature regime, $\delta \omega_{\rm SD}$ becomes smaller,
indicating that the spin-dipole oscillation can qualitatively be explained by the mean-field theory and
pairing fluctuations are suppressed by the appearance of the superfluid order in this regime.
\par
\section{Summary}
\label{sec4}
To summarize, we have studied the spin-dipole frequency in a trapped Fermi gas near the unitarity by using the extended $T$-matrix approximation combined with the local density approximation and the sum rule approach.
We have showed that our numerical result is in an excellent agreement with the recent experimental result at the unitarity limit.
In the classical (high-temperature, strong coupling) regime, the analytical expression for the spin-dipole frequency has been derived.

In particular, the calculated spin-dipole frequency exhibits a large enhancement in the low-temperature regime, due to the formation of the spin-singlet pairs in the center of the trap, i.e., due to the formation of a sizable region of zero magnetic susceptibility.
The spin-dipole frequency coincides with the trap frequency in the high-temperature regime of the BCS-unitary region, in accordance with Kohn's theorem. 
In the strong-coupling regime, where a molecular bound state is present,
the spin-dipole frequency deviates from the trap frequency in the region where $T\lesssim E_{\rm b}$.
%We extracted the frequency shift of the spin-dipole oscillation originating from the pairing correlations beyond the mean-field approximation.
%This quantity has a maximum just below $T_{\rm c}$ ($T=0.77T_{\rm c}$), where the system is dominated by strong pairing fluctuations.
%In the low-temperature limit, the divergence of the spin-dipole frequency is related to the emergence of the mean-field superfluid order parameter.
\par
While this work focused on the mass-balanced case, our results could be generalized to mass-imbalanced systems. 
This is relevant for the new generation of Fermi-Fermi mixture experiment where the overlap of the two Fermi clouds can be large as in the case of
disprosium-potassium mixtures~\cite{DyK-Rudi,DyK-Rudi2}. 
Regarding this, we comment on the fact that the upper bound Eq. (\ref{eq12}) could even be improved 
by considering that the low-energy quasi-particle excitations do not exhaust the f-sum rule $m_1$.
While, for our equal mass case, the correction would be rather small, it could be an interesting future problem in mass-imbalanced mixtures.

%\begin{acknowledgment}
\acknowledgments
H. T. thanks K. Iida, F. Scazza, R. Hanai, M. Pini, and M. Ota for useful discussions.
This work was supported by a Grant-in-Aid for JSPS fellows (No.17J03975).
A. R. acknowledge support from the Provincia Autonoma di Trento.
Y. O. was supported by a Grant-in-Aid for Scientific Research from MEXT and JSPS in Japan (No.JP18K11345, No.JP18H05406, and No.JP19K03689).
%\end{acknowledgment}

%\appendix
%\section{}

%Use the \verb|\appendix| command if you need an appendix(es). The \verb|\section| command should follow even though there is no title for the appendix (see above in the source of this file).

%For authors of Invited Review Papers, the \verb|profile| command is prepared for the author(s)' profile.  A simple example is shown below.

%\begin{verbatim}
%\profile{Taro Butsuri}{was born in Tokyo, Japan in 1965. ...}
%\end{verbatim}

\end{document}